\documentclass[conference]{IEEEtran}
\IEEEoverridecommandlockouts
\usepackage{cite}
\usepackage{amsmath,amssymb,amsfonts}
\usepackage{algorithmic}
\usepackage{graphicx}
\usepackage{textcomp}
\usepackage{xcolor}
\usepackage{epstopdf}
\def\BibTeX{{\rm B\kern-.05em{\sc i\kern-.025em b}\kern-.08em
    T\kern-.1667em\lower.7ex\hbox{E}\kern-.125emX}}

\newcommand{\Cb}{{\emph{Cynopterus brachyotis}}}
\newcommand{\Gs}{{\emph{Glossophaga soricina}}}
    
\begin{document}

\author{Xiaozhou Fan$^{1}$, Alexander Gehrke, and Kenneth Breuer\\

Center for Fluid Mechanics, School of Engineering, Brown University, Providence, RI 02912 USA

\thanks{$^{1}$Correspondent author. Current position: Postdoc at GALCIT, Caltech, Pasadena CA 91106 USA}
}

\title{Wing twist and folding work in synergy to propel flapping wing animals and robots}


\maketitle

\begin{abstract} 
We designed and built a three degrees-of-freedom (DOF) flapping wing robot, \textit{Flapperoo}, to study the aerodynamic benefits of wing folding and twisting. Forces and moments of this physical model are measured in wind tunnel tests over a Strouhal number range of $St = 0.2 – 0.4$ -  typical for animal flight. We perform particle image velocimetry (PIV) measurements to visualize the air jet produced by wing clapping under the ventral side of the body when wing folding is at the extreme. The results show that this jet can be directed by controlling the wing twist at the moment of clapping, which leads to greatly enhanced cycle-averaged thrust, especially at high $St$ or low flight speeds. Additional benefits of more thrust and less negative lift are gained during upstroke using wing twist. Remarkably, less total actuating force, or less total power, is required during upstroke with wing twist. These findings emphasize the benefits of critical wing articulation for the future flapping wing/fin robots and for an accurate test platform to study natural flapping wing flight or underwater vehicles.

\end{abstract}

\begin{IEEEkeywords}
flapping wing, robotics, bio-inspired propulsion
\end{IEEEkeywords}

\section{Introduction} 

\begin{figure}
    \vspace*{-20pt}
    \centering
    \includegraphics[width=0.5\textwidth]{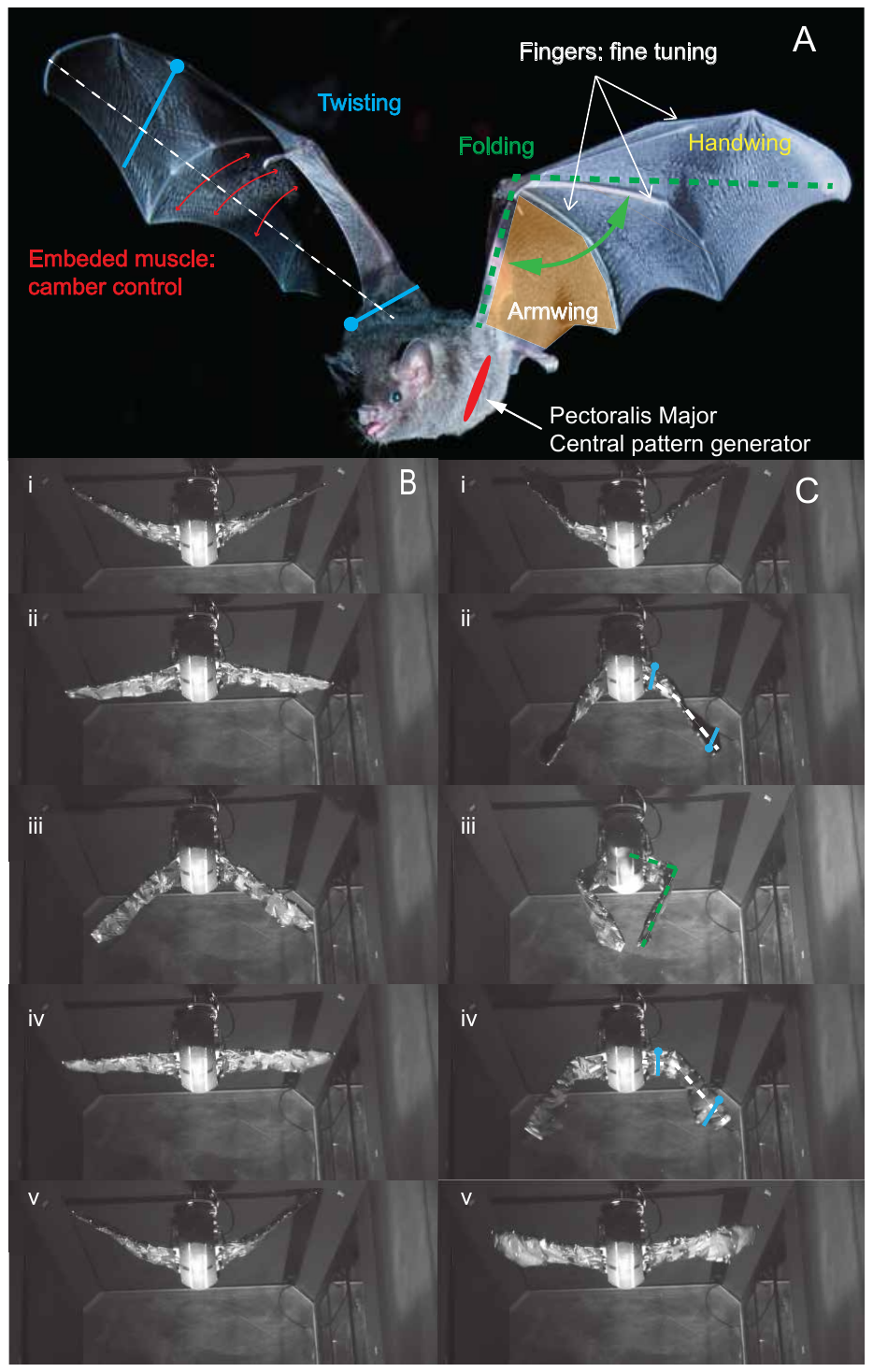}
    \vspace*{-20pt}
    \caption{Wing twist and folding in animal flight and Flapperoo - a three-degree-of-freedom robotic platform capable of wing flapping, twist and folding. (A) A fruigivorous bat, \Gs~, in flight (photo credit: Brock Fenton). Wing twist refers to the variation of pitching angles along the wingspan, as indicated by the blue sticks (filled circles represent the leading edges). Wing folding is the additional rotation of outboard wing with respect to the inboard wing. (B) Without folding and twist, the wings simply beat up and down. (C) The wings of Flapperoo fold inwards from late downstroke to middle upstroke (green dashed), and unfold for the remaining upstroke. Due to active wing twist, the outboard wing has a nose-down pitch during the downstroke (differences in orientation of blue sticks), while it pitches back up during upstroke.}
    \label{fig:intro}
\end{figure}

Flapping wing flight offers prominent advantages over rotorcraft and fixed-wing drones, such as improved flight efficiency, greater lift production \cite{Wang2004}, and lower acoustic signatures at low Reynolds numbers ($10^2 \sim 10^5$) \cite{Jaworski2020}.
However, the challenges are equally evident. For example, the inseparable coupling between unsteady aerodynamics (involving complicated wing-wing interaction as found in dragonflies \cite{Hefler2020,Noda2023}), flight control (both passive and active \cite{Liu2022}), and the onboard power electronics/driving mechanism. These factors need to be carefully considered together in designing of flapping wing robots \cite{Shyy2016}. 

On the other hand, robotic flapping wing platforms lend itself to tease apart the impact of wing kinematics on aerodynamics, as they closely and consistently mimic the animal flight in a controlled environment.
Live animal experiments have shed light on flight mechanics for bats and birds \cite{Riskin2008,Hedrick2004,Windes2018}, however, in certain scenarios, such as the investigation of jet propulsion due to ventral wing clapping \cite{Chang2011,Chang2013}, animal experiments may still be challenging to perform in a consistent and controlled manner \cite{Hedrick2004}.
Indeed, studies show that bats control dozens of their wing joints actuated by muscles and tendons (Fig.~\ref{fig:intro}A) \cite{Norberg1972,Bahlman2013,Bender2019,Fan2018}, which renders direct replication in a robotic system extremely difficult.
On the other hand, some degrees-of-freedom (DOF) of the wings may not be relevant to flight performance, but are rather dedicated to the animal's everyday activity, such as foraging or mating \cite{Norberg1972,Riskin2008}.

Birds and bats (loosely refer to animals of wingspan $>40$cm, with Reynolds number $10^4 - 10^5$) modulate their wing kinematics, such as wing twist and folding angles, within each cycle and across different flight speeds \cite{Parslew2015,Fan2022}.
The pectoralis major muscle, as identified by Biewener \textit{et al.} \cite{Biewener1998}, serves as a central pattern generator (CPG, Fig.~\ref{fig:intro}A). As discussed by Marder \textit{et al.} \cite{Marder2001} for a wide range of animals, the armwing (or inboard wing, driven by CPG) of bat remains relatively insensitive to changes in flight speed - with fixed flapping frequency and armwing wingbeat amplitudes, except for camber control that does change with speed \cite{Fan2022}.
In contrast, the distal handwing (outboard wing), which is lightweight and capable of adapting to speed variations, emerges as an optimal control surface for generating more varied aerodynamic forces \cite{Fan2022}.

Robots that only consider passive articulation and/or alter flapping frequency to modulate force generation often have a lower flight efficiency. Wissa \textit{et al.} \cite{Wissa2012} demonstrated that the passive wing articulation at this bird-scale can be inefficient, with a power consumption reaching $60 - 80$W for flapping frequency of $4-6$Hz. Bie \textit{et al.} \cite{Bie2021} added a passive folding mechanism to the same four-bar linkage mechanism. Their flight time was also rather short ($\sim$ 5 min).
On the other hand, Ramezani \textit{et al.} \cite{Ramezani2017} introduced a bat-inspired robot (B2) with active wing retraction, and demonstrated promising flight capabilities. Perhaps a major breakthrough in achieving low flight power requirements came from Festo’s Smartbird by Send \textit{et al.} \cite{send2012}, which performed a fixed amount of wing folding and was capable of active twist using a servo motor. These active wing DOF dramatically reduced consumption to around $20$W. However, in their design, the wing folding was actuated by a fixed four-bar linkage, in contrast with bats and birds who actively adjust the wing folding angle with flight speed \cite{Hubel2016,Parslew2015}. Moreover, the wing twist of Smartbird was controlled in binary states - \textit{i.e.}, either a fixed positive or negative angle of incident depending on it is up or downstroke - as opposed to a time-varying function within a cycle. This constraint would likely prevent the wings from achieving optimal local effective angle of attack \cite{Parslew2012,Sekhar2018}. Lastly, the servos that control the twist were mounted near the wing tip, which increases the moment of inertia of the wing, and thereby increases the inertial power required to flap the wing \cite{Riskin2012}.

While active wing articulation proves important, but to quantitatively home in on the ensuing aerodynamics, stationary-mounted flapping wing platforms, either in wind or water tunnel environments, are necessary steps. These platforms offer the ability for closer mimicry of the observed animal wing motions since the weight of the robot is no longer a critical concern. Inspired by the frugivorous lesser-nosed dog-faced bat, \Cb , Bahlman \textit{et al.} \cite{Bahlman2013} built a robotic wing and subsequently studied the cost of flight using this platform \cite{Bahlman2014}. 
Chen \textit{et al.} \cite{Chen2021} designed a 10 DOF robot (5 DOF for each wing) that was inspired by Passerine. Hovering in a water tank testing facility, they measured the lift forces as a result of observed wing kinematics (flapping, sweeping, twisting, folding) and found wing folding helps mitigate negative lift during upstrokes. It is noted, however, that the measured forces would only explain 25\% of the target animal's weight.

It is clear that many open questions in unsteady aerodynamics/propulsion still remain. Here we present a robust three DOF flapping wing platform   - ``Flapperoo'' - that showcases the synergistic wing motion of twisting and folding observed in bat flight (Fig.~\ref{fig:intro}C), as motivated by recent reduced-order dynamical system modeling of bat flight\cite{fan2021a,Fan2021b}. Here, we present the detailed design that enables wing twist and folding motions to be fully adaptive to flight speed - both by magnitude and manner of actuation, which offers a vast parameter space. 
These two wing motions, when combined, offer unconventional bio-inspired propulsion by means of wing clapping on the ventral side of Flapperoo during upstroke, with which the jet direction can be manipulated by means of wing twist.

\section{Methods}

\begin{figure*}
    \centering
    \includegraphics[width=\textwidth]{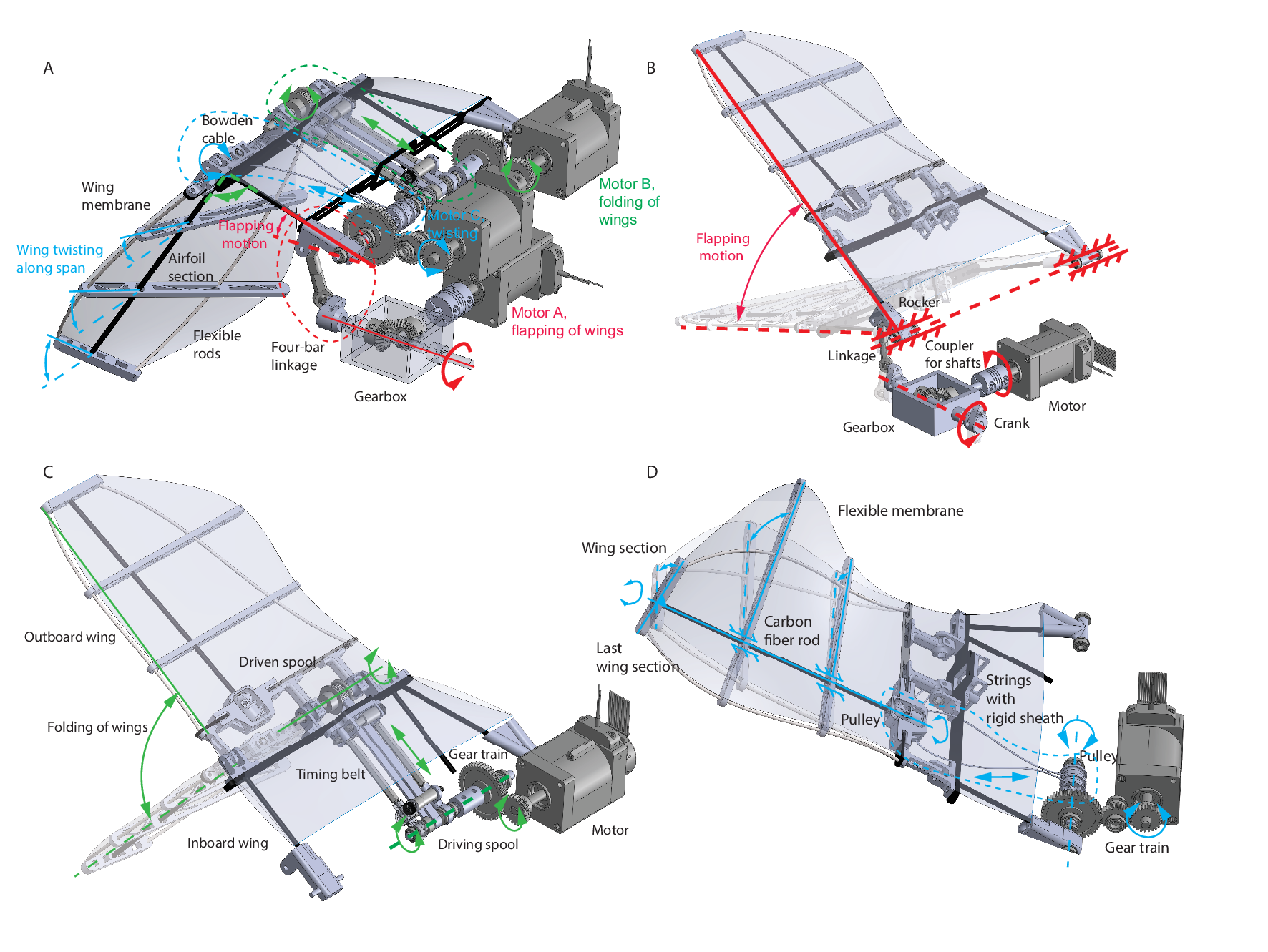}
    \caption{Design of the driving mechanism and wing articulation. (A) Close-up view of the flapping, folding and wing twist design. 
    (B) The flapping motion (red) is driven by a four-bar linkage mechanism. (C) Wing folding mechanism schematics. (D) Wing twist actuation schematics.}
    \label{fig:mechDesign}
\end{figure*}

\textit{Flapperoo} is a bat-inspired robotic platform that can independently control its flapping, wing folding, and wing twist motion. The choice of these two active wing motions stems from live bat experimental and simulation studies \cite{Riskin2008,fan2021a,Fan2022} and mathematical modeling of the wing-body dynamical system optimized on the power consumption \cite{Fan2021b}. The wingspan of Flapperoo is $\sim 70$ cm, and the averaged chord length is $15$ cm. An aerodynamically smooth body houses the mechanism to minimize flow separation.


The primary components of the flapping mechanism are fabricated from 3D printed plastic and carbon fiber. As depicted in Fig.~\ref{fig:mechDesign}A, the primary flapping motion (text colored in red) is driven by a four-bar linkage mechanism. The twisting motion (in blue) is actuated by a Bowden cable mechanism. The wing folding (in green text) is realized using a timing belt transmission, where the outboard portion of the wing is fixed with a spool that rotates with respect to the inner portion of the wing. 

Specifically, for the design of the flapping motion in Fig.~\ref{fig:mechDesign}B, the motor spins continuously in one direction, distributing power to both left and right wing using a bevel gear. This drives a crank, connected to a  linkage (or coupler) which in turn is connected to a rocker that flaps the wing. As outlined in Fig.~\ref{fig:mechDesign}C, to drive the wing folding, a second motor outputs a reciprocal motion that rotates a driving spool and timing belt back and forth. At the other end of the belt, the driven spool sits at the junction between the inner and outer part of the wing, and thereby producing the folding motion.
A third motor (Fig.~\ref{fig:mechDesign}D), which also rotates reciprocally,  is used to achieve the wing twist using antagonistically-tensioned cables that alternately pull and release, causing the last distal wing section to rotate. The wing sections are linked using flexible rods, generating a gradual reduction in pitching angle that is is effectively manifested as wing twist.  We use a flexible, in-extensible membrane to cover the wing surface.
The wing and all exposed parts are painted in matte black to avoid reflections during the particle image velocimetry (PIV) experiments.

We use a motion controller (DMC-4040, Galil Motion Control, USA) to drive and control the three motors (BE163CJ-NFON, Parker Hannifin Corp., Rohnert Park, CA) for the flapping, folding, and twist actuation.

A dynamic model of the three wing angles and their interaction, is used to decouple the fold and twist with respect to the flapping angle position, and thus control the three DOF wing motion with complete generality. 
A custom Matlab script \cite{Gehrke2021} is used to coordinate the entire motion.

    \begin{figure}
        \centering
        \includegraphics[width=0.4\textwidth]{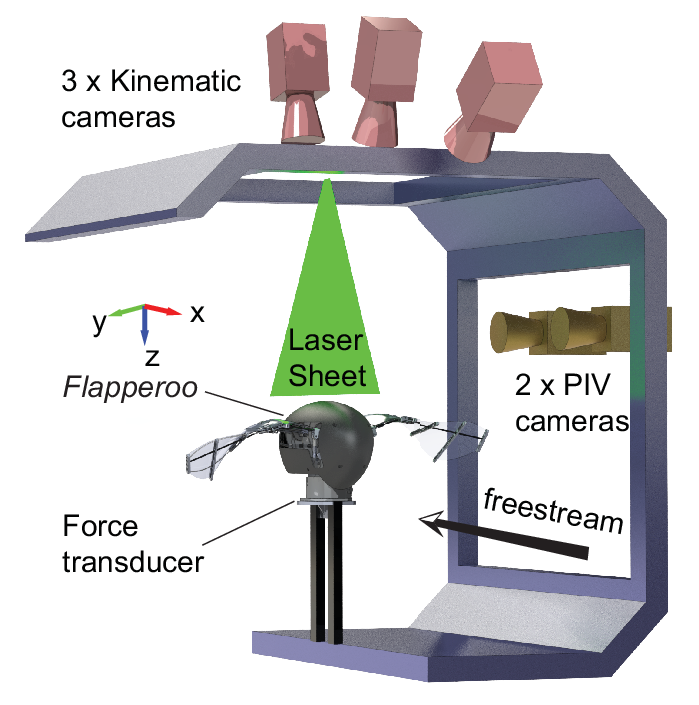}
        \caption{The robot was tested in a closed-loop low turbulence level wind tunnel at Brown University \cite{Breuer2022}. }
        \label{fig:expr setup}
    \end{figure}
    
During testing, we set the flapping frequency at $f =  3$Hz, and varied the freestream velocity to achieve different Strouhal numbers, $St = fA/U$, where $A$ is the vertical wingtip displacement when there is no folding, and $U$ is the freestream velocity \cite{Taylor2003,Riskin2008}. Flapperoo was mounted upside down in the wind tunnel on a six-axis force/torque sensor (Gamma IP65, ATI Industrial Automation, NC) and force measurements were recorded using an A/D converter (USB-6343, National Instrument, TX) at $1000$Hz (Fig.~\ref{fig:expr setup}). The $x$-axis points forward to the streamwise direction, and the $z$-axis is vertically downward. The pure inertial cost of the system was carefully subtracted to separate the aerodynamic and inertial contributions. The angular position of the motors was recorded at $512$Hz using the Galil motion controller. White reflective markers, placed at the leading edges of the wings, were tracked  at 300 fps using three high speed cameras (Phantom Miro 340 Vision Research Inc.). The video and force data acquisition were synchronized using an Arduino microcontroller. The cameras were calibrated and the videos were digitized using the video digitizing and annotation tool DLTdv8 \cite{Hedrick2008}. During the particle image velocimetry (PIV) experiments a Nd:YLF double-pulsed laser (DM40, Photonics Industries, Ronkonkoma, NY) was employed at $300$Hz with an energy output of approximately $40$ mJ/pulse. The vertical laser sheet was aligned with the flow direction and located at the body's midline which coincides with the center of the gap between the wingtips almost touch to clap. The test section was seeded with DEHS tracer particles (approx 1 micron in diameter) which were imaged using two FASTCAM NOVA R2 high-speed cameras (2048 x 2048 pixels, Photron USA, Inc.), positioned side by side with 50\% overlapping fields of view, producing a combined field of view of $450 \times 350$mm. DaVis PIV software v10 (LaVision Inc., Germany) was used to determine the resulting flow-field.

    \begin{figure}
        \centering
        \includegraphics[width=0.4\textwidth]{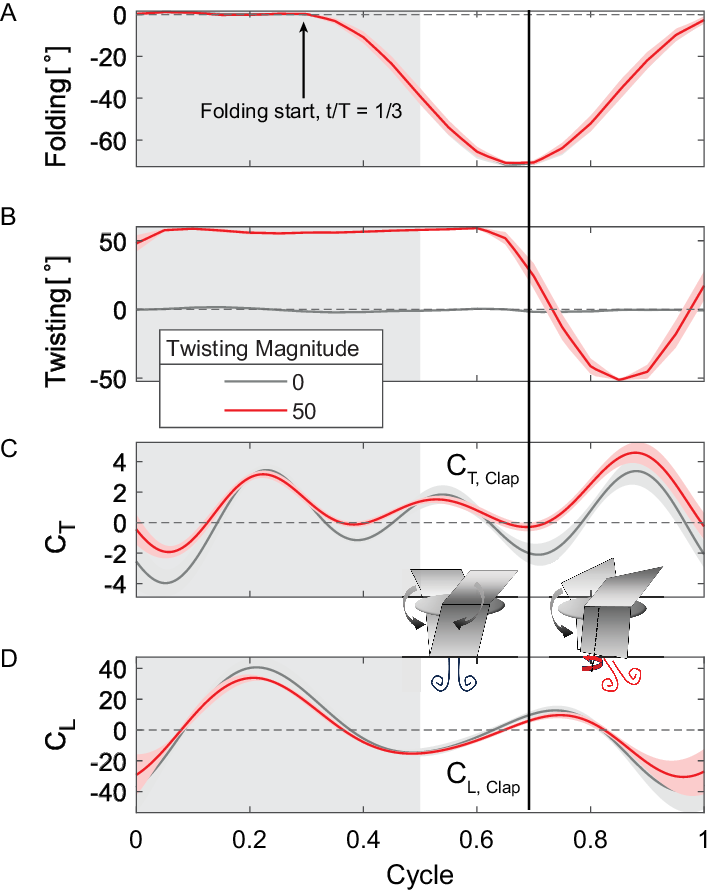}
        \caption{Time series of the wing folding and twisting actuation, thrust and lift measurements. Note the folding is already at maximum for twisting or non-twisting cases, and the wings would clap at the ventral side. The shaded area indicates the downstroke. The standard deviation calculated over 50 cycles is indicated by the bands around the signal. The black vertically running line indicates the moment of wing clapping, and the two cartoon insets illustrate the air jet produced and that is directed or ``vectorized'' with the wing twist.}
        \label{fig:kineLiftThrust}
    \end{figure}
%
Two cases, one with no wing twist and one with maximum wing twist, are presented in Fig.~\ref{fig:kineLiftThrust}A. Note that both cases have a folding amplitude of $70^\circ$, where the wings clap together at around $t/T = 0.7$ (indicated by the black vertical line). The down- and upstroke are defined by the motion of the inboard wing, spanning exactly half of a cycle each (shaded vs. white area in Fig.~\ref{fig:kineLiftThrust}). The start of the folding is programmed to occur late into the downstroke, around $t/T = 1/3$. For the non-twisting case (grey line in Fig.~\ref{fig:kineLiftThrust}B), the twist stays effectively zero throughout the entire cycle, whereas the case with actively articulated twist, the wing tip during the downstroke begins with and holds a nose-down pitching of $50^\circ$ throughout the entire downstroke. Moving inboard, the wing pitch angle decreases reaching zero at the wrist. At the time of the wing clap, $t/T = 0.7$, the wings supinate, and their wing tips rotate rapidly and transit from a nose-down to nose-up pitch. The pitch angle reaches $-50^\circ$ towards the end of upstroke ($t/T = 0.85$). The wing pronates around $t/T = 0.9$ before the next downstroke begins.

\section{Results and discussion} 
Design of Flapperoo is deeply rooted in the flight mechanics of birds and bats - it has a speed invariant armwing (inboard wing) that is driven a set four-bar linkage, and an adaptive and light handwing (outboard wing) that changes with flight speed. All the heavy components - including motors, gears, shafts are located as close to the body as possible to minimize the inertia penalty in flapping - which is also anatomically accurate weight distribution for birds \cite{Hedrick2004} and bats \cite{Fan2022}.

The experiments were designed to study 1) how the air jet, produced by the wing clap during upstroke on the ventral side of the body, may be redirected using active wing twisting, 2) how the propulsive force scales with biologically relevant parameters such as the Strouhal number $St$, and finally 3) how wing twist may contribute to more thrust and positive lift during upstroke, with even smaller total force (less power requirement). 

Note in the experiment, since \textit{Flapperoo} is mounted upside down, thus the positive lift direction is physically vertically downward as shown in Fig.~\ref{fig:expr setup}.

\subsection{Direction of air jet}


In Fig.~\ref{fig:kineLiftThrust}C and D, we compare the time series of recorded forces for wing twist and non-twist cases. The freestream velocity in these cases is $U_\infty = 2$ m/s ($St = 0.42$). For the non-dimensional thrust force coefficient in the streamwise direction, normalized as $C_T = F_x/0.5 \rho U_\infty^2 A$ ($A$ is the wing area), the twisting case consistently generates more positive thrust throughout the cycle (Fig.~\ref{fig:kineLiftThrust}C). In the vertical direction, the lift coefficient $C_L= F_x/0.5 \rho U_\infty^2 A$, and the twisting case slightly produces less positive lift except for late upstroke.

\begin{figure}
    \centering
    \includegraphics[width=0.45\textwidth]{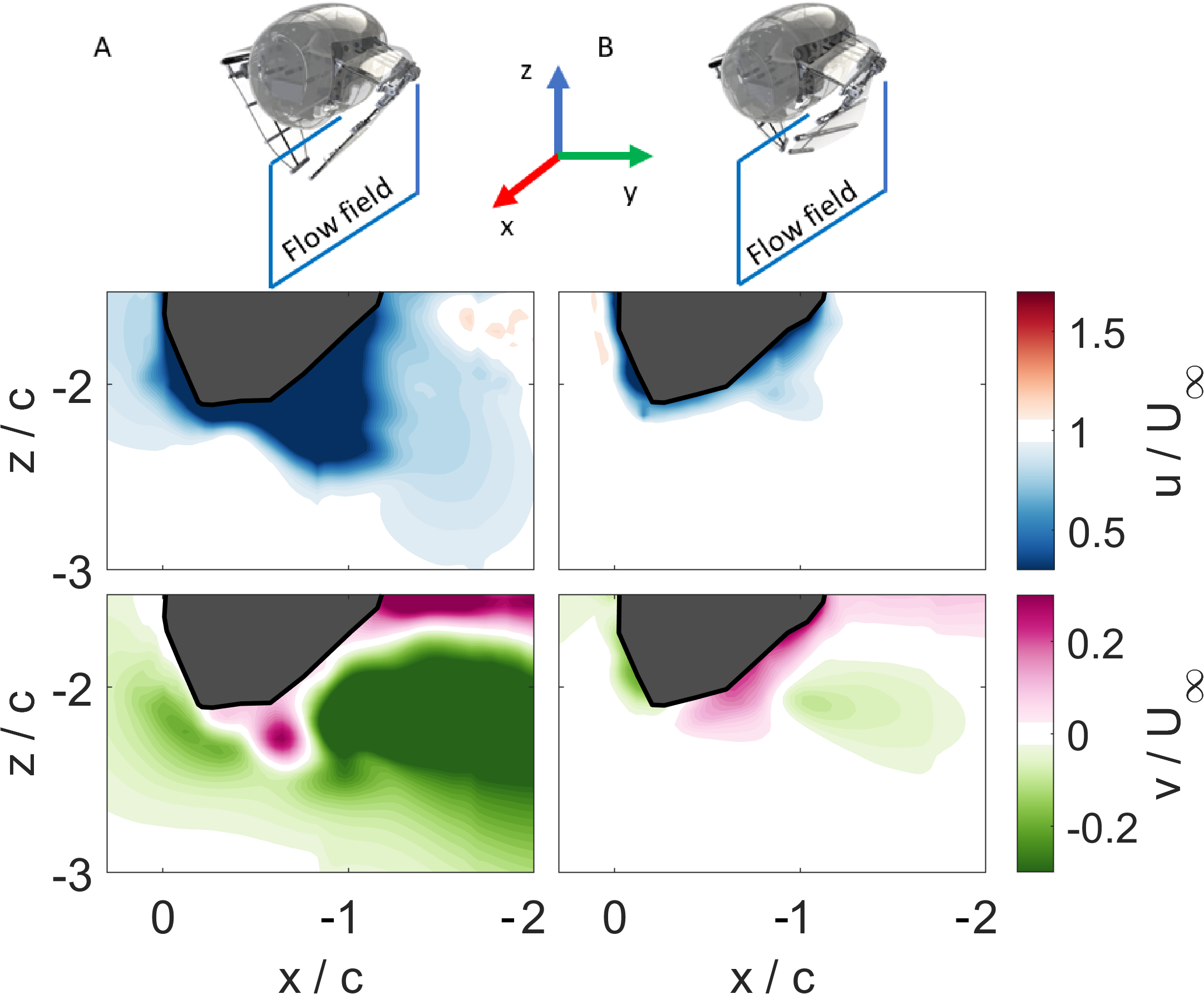}
    \caption{Comparison of streamwise $u$ and vertical $v$ velocity, normalized by the oncoming flow speed, $U_\infty$, at the moment of wing clapping $(t/T = 0.7)$ for no twist (A) and maximum wing twist (B). $U_{\infty} = 4$m/s or $St = 0.21$.}
    \label{fig:PIV}
\end{figure}

To link the aerodynamic force production to the redirection of the clapping air jet (propulsion vectoring), we present the streamwise, $u/U_{\infty}$, and vertical, $v/U_{\infty}$, velocity components at moment of clap $(t/T = 0.7)$ in Fig.~\ref{fig:PIV}. In the streamwise direction, without wing twist, a large region of fluid is slowed down around the wing during the clap.
In contrast, with wing twist, the region of flow retardation is greatly reduced.
Similarly, we see in the vertical velocity, wings with twist, produces a much weaker region of downward moving fluid compared to the zero-twist case.
The low-speed region behind the wing is associated with a drag force on the wing (black vertical line at $t/T = 0.7$, Fig.~\ref{fig:kineLiftThrust}C), while the downward-directed fluid means the additional lift is being generated ($t/T = 0.7$, Fig.~\ref{fig:kineLiftThrust}D).


    \begin{figure}
        \centering
        \includegraphics[width=0.4\textwidth]{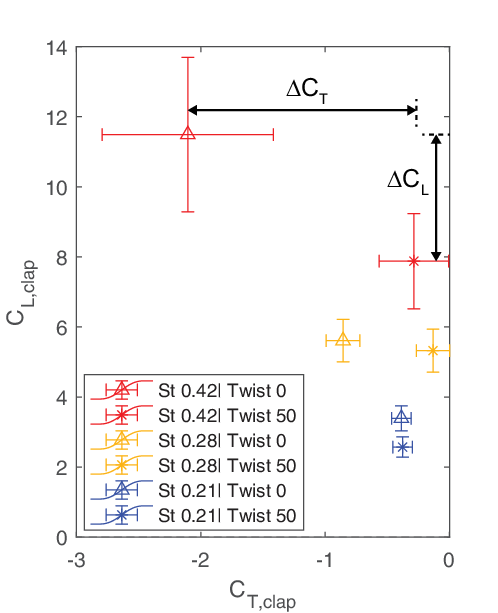}
        \caption{Thrust and lift force coefficients, $C_{T,clap}$ and $C_{L,clap}$ respectively, at the moment of the wing clapping ($t/T = 0.7$ in Fig.~\ref{fig:kineLiftThrust}) for twisting and non-twisting and a range of different $St$. The errorbars are the standard deviation from 50 cycles.} 
        \label{fig:clapCmp}
    \end{figure}

In order to understand how Strouhal number $St$ may scale the propulsive force, we compare the changes in $C_{L,clap}$ and $C_{T,clap}$ for the non-twist and twist cases over a range of $St$ (Fig.~\ref{fig:clapCmp}). These instantaneous coefficients are taken right when the wings would clap at $t/T = 0.7$ (vertical line in Fig.~\ref{fig:kineLiftThrust}). We find that clapping the wings together at an angle (\textit{i.e.} with wing twist) increases thrust $\Delta C_T$, but at a cost of decreased lift, $\Delta C_L$. This trade-off between thrust and lift production is strongly dependent on $St$. The gain $\Delta C_T$ is minimal at low $St$, but becomes significant at high $St$. The drop in lift, $\Delta C_L$, is most apparent at the two extremes of $St = 0.21$ and $0.42$ and less significant at $St \sim 0.28$, perhaps suggesting that $St \sim 0.28$ might be a suitable region where more thrust can be produced with little compromise in lift from wing clapping and twisting.

Note, at $St = 0.21$, the PIV experiment suggested the flow behind the wing is slowed down much more significantly for case without wing twisting (streamwise velocity field $u/U_\infty$ in Fig.~\ref{fig:PIV}), but the drag recorded by force transducer is not increased (Fig.~\ref{fig:clapCmp}). This is due to the projected wing area $A$ in the $z-y$ plane is much smaller for the non-twisting case when compared to the twisting case. Thus the actual drag Flapperoo experienced is also small. On the other hand, for the vertical velocity component, $v$, the horizontal projected area in $x-y$ plane for both cases are comparable, and thus the acceleration of the downward fluid flow agrees well with the net ``jump'' in lift coefficient in Fig.~\ref{fig:clapCmp}.

\begin{figure}
    \centering
    \includegraphics[width=0.4\textwidth]{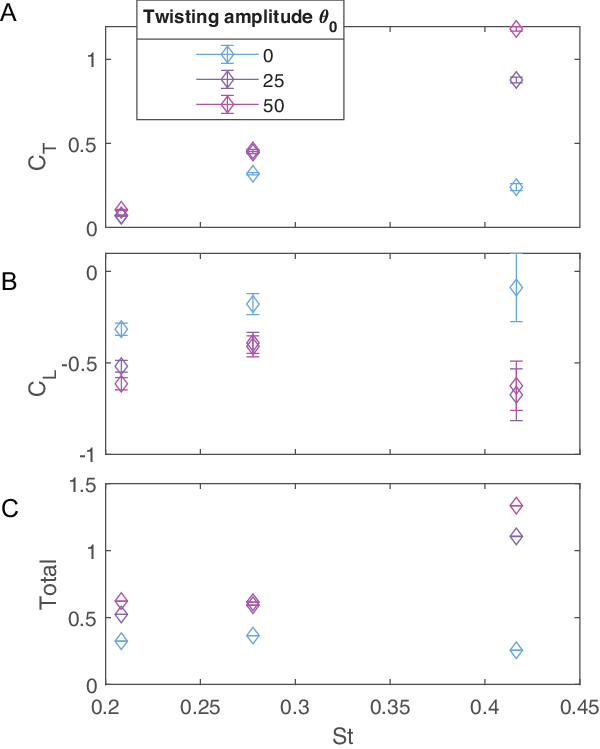}
    \caption{Scaling of the cycle-averaged thrust coefficient as a function of Strouhal number for different twisting amplitudes. All cases have maximum folding angles of $70^\circ$ where the wings clap. The errorbar are the standard deviation of 50 cycles.} 
    \label{fig:scaling}
\end{figure}

Wing twist leads to an increase in thrust during the moment of clap, as well as the overall cycle-averaged force production (Fig.~\ref{fig:scaling}). Without wing twist, no clear trend of the cycle-averaged thrust can be observed as a function of Strouhal number $St$ (Fig.~\ref{fig:scaling}A). However, with wing twist, cycle-averaged thrust increases monotonically with increasing $St$. For a fixed $St$, more twist yields generally higher cycle-averaged thrust, $C_T$, which is most pronounced at higher Strouhal numbers. On the other hand, the cycle-averaged lift, $C_L$, decreases with increasing twisting amplitude for a fixed $St$ (Fig.~\ref{fig:scaling}B). It is interesting to note that when there is no twisting, $St$ does  provide a good scaling for $C_L$. 
Without wing twist, the total force,$\sqrt{(C_L)^2+(C_T)^2}$, is independent of $St$ (Fig.~\ref{fig:scaling}C). With wing twist, however, the total force increases monotonically with $St$.
For a fixed $St$, more twisting yields a higher overall force production. 
In summary, without twist, $St$ scales $C_L$ well, but when there is wing twist, $St$ scales $C_T$ instead.

Finally, it should be emphasized that the wing kinematics chosen in this work are not necessarily optimal. For example, the absolute values of the cycle-averaged lift ($C_L$ in Fig.~\ref{fig:scaling}B) are all close to or even below zero, which suggests that the timing and magnitude of wing folding and twist might be further optimized to yield improved aerodynamic performance. Nevertheless, wing twisting offers flapping wing animals/robots another symmetry-breaking mechanism, in the streamwise $x-$axis, and enables the directional control of the bio-inspired jet propulsion in the form of wing clapping.

\subsection{Effect of wing twist during the upstroke}

\begin{figure}
    \centering
    \includegraphics[width=0.4\textwidth]{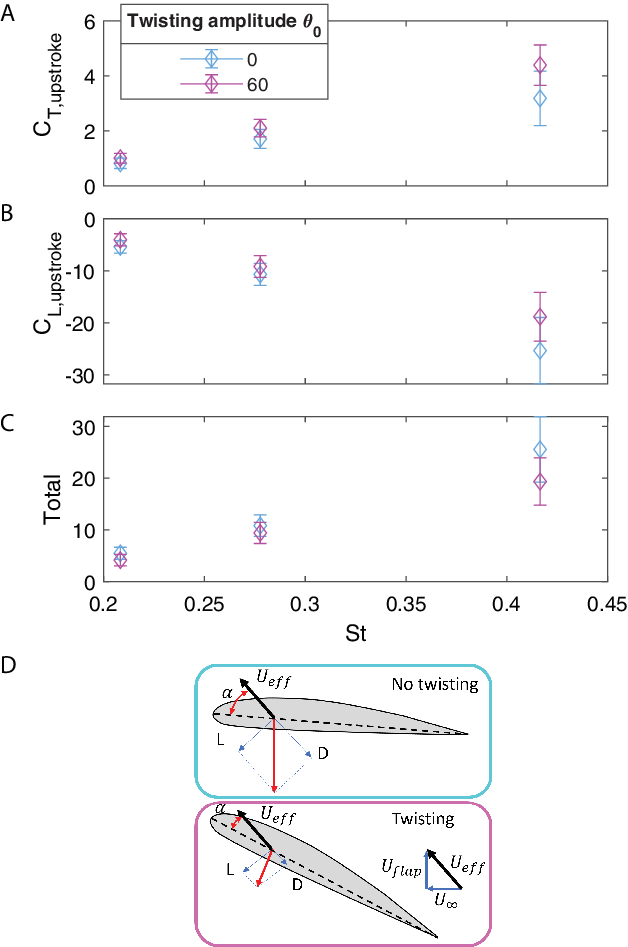}
    \caption{Upstroke force production with twisting and without twisting comparison, around snapshots $t/T = 0.9$ in Fig.~\ref{fig:intro}B(v) and C(v). (A) Thrust coefficient $C_{T,clap}$. (B) Lift coefficient $C_{T,clap}$ (C) Total force generation, $\sqrt{(C_L)^2+(C_T)^2}$. The errorbar are the standard deviation of 50 cycles. (D) A cartoon to explain the underlying mechanics in quasi-steady arguments.  } 
    \label{fig:upstrokeForce}
\end{figure}

The synergy between wing folding and twist is also demonstrated during the late upstroke, when the wings move ventral-dorsally just before the next downstroke. More thrust and less negative lift is being generated during this recovery stroke for the wing twist case (around $t/T = 0.9$ in Fig.~\ref{fig:kineLiftThrust}C and D), and this trend holds true across the entire range of $St = 0.2 - 0.4$ (Fig.~\ref{fig:upstrokeForce}A and B). Specifically, more thrust is being generated during upstroke for the case with wing twist, while fewer losses in lift are experienced. This trend increases with $St$, and leads to overall less total forces produced for cases with wing twist during upstroke; the difference further increases with $St$ (Fig.~\ref{fig:upstrokeForce}C). Because the total force directly relates to the required actuating force at the shoulder joint,  this suggests a lower power requirement during the upstroke. We can use a quasi-steady argument to understand the underlying mechanics (Fig.~\ref{fig:upstrokeForce}D) - holding other conditions constant, wing twist reduces the effective angle of attack, $\alpha$, which reduces both lift and drag as produced by the wing. However, drag experiences a much steeper reduction than lift (Fig. 2C in Dickinson \textit{et al.} \cite{Dickinson1999}) 
and as a result, the total force vector not only shrinks, but also orients more towards the flight direction. This results in  ``one stone, three birds'' -  more favorable thrust, less negative lift and lower power requirement.

\section{Conclusion/Summary}

A bat-inspired three DOF flapping wing robot, Flapperoo, capable of adaptive wing twist and folding has been designed, built and tested. We show that these two wing motions enable a ``vectorized'' jet propulsion during upstroke by means of ventral wing clapping. More favorable forces (thrust and positive lift) are generated through the synergy at a reduced cost in actuating force or power required. This platform lends itself to an optimal kinematics search, which is impossible in live animal experiments. These findings are also critical for progress in developing flapping wing robots where flight efficiency and control authority is important. 

\section*{Conflict of interest}
A US non-provisional patent No. $18/514,625$ is filed for the design of flapping wing platform.

\section*{Acknowledgments}
Gratitude goes to Dr. Alberto Bortoni for his kind hand in setting up the motion tracking system, and Siyang Hao and R\'on\'an Gissler for their time and assistance to help with the PIV experiments. We would also like to thank Amick Sollenberger, who crafted an aerodynamic body for Flapperoo, and to Kaela Zaino's kind feedback on this manuscript. Finally, much gratitude goes to Dr. Karen Mulleners and her lab, for her generous discussions on using Matlab functions to drive motors. This work is supported by the US Office of Naval Research, Award N00014-21-1-2816.  AG is also supported by the US National Science Foundation, Award CBET-2035002. 

\bibliographystyle{ieeetr}
\bibliography{root}


\end{document}